\documentstyle[12pt,aaspp4]{article}
\input{epsf}
\begin{document}
\def\msun{\hbox{${\cal{M}}_{\odot}$}}
\def\massA{\hbox{${\cal{M}}_A$}}
\def\massB{\hbox{${\cal{M}}_B$}}
\def\massAB{\hbox{${\cal{M}}_{A+B}$}}

\title{Speckle Interferometry at the U.S. Naval Observatory. XXII.}

\author{Brian D.\ Mason and William I.\ Hartkopf}
\affil{U.S. Naval Observatory \\
3450 Massachusetts Avenue, NW, Washington, DC, 20392-5420 \\
Electronic mail: (brian.d.mason, william.hartkopf)@navy.mil}

\begin{abstract}

The results of 4,747 intensified CCD observations of double stars, 
made with the 26-inch refractor of the U.S. Naval Observatory, are 
presented. Each observation of a system represents a combination of 
over two thousand short-exposure images. These observations are 
averaged into 2,667 mean relative positions and range in separation 
from 0\farcs328 to 95\farcs9, with a median separation of 
8\farcs673. Four orbits are improved. This is the 22$^{nd}$ in this
series of papers and covers the period 4 January 2016 through 29 
December 2016. 
\end{abstract}

\keywords{binaries : general --- binaries : visual --- stars : 
individual(BD$+$32\phn1731, HD 183063, HD 193651) --- techniques :
interferometric}

\section{Introduction}

This is the 22$^{nd}$ in a series of papers from the U.S. Naval 
Observatory's speckle interferometry program, presenting results of 
observations obtained at the USNO 26-inch telescope in Washington, 
DC (see, most recently, Mason et al.\ 2017). 

From 4 January through 29 December 2016, the 26-inch telescope 
was used on 67 of 216 (31\%) scheduled nights. While most nights 
were lost due to weather conditions, time was also lost due to 
testing and upgrades of instrumentation and software, other 
mechanical or software issues, and to a lack of observing personnel.

Individual nightly totals varied substantially, from 7 to 200 
observations per night (mean 71.3). The results yielded 4747
observations (pointings of the telescope) and 4795 resolutions 
(pairs measured). Not all observations result in a resolution and 
observations of multiple star systems in a single CCD field gives 
the somewhat non-intuitive result where the number of resolutions 
exceeds the number of observations. After removing marginal 
observations, calibration data, tests, and $``$questionable 
measures" a total of 4143 measurements remained. These 
$``$questionable measures" are not all of inferior quality but may 
represent significant differences from the last measure, often made
many decades ago. Before these measures are published they will 
need to be confirmed in a new observing season to account for any 
possible pointing or other identification problems. The tabulated 
list of these is retained internally and forms a $``$high priority 
observing list" for subsequent observing seasons. In this 
continuing process we here also present 38 measures obtained from 
2002-2005. These 4181 measures were grouped into 2667 mean 
relative positions. 

\section{Instrumentation \& Observing Technique}

In \S2.3 of Mason et al.\ (2017) we described future planned 
improvements of the USNO visual double star observing system. 
Specifically, item \#1, which was illustrated in Figure 1, allows
us to observe closer pairs. This has several important consequences.
Rather than using double stars for scale and rotation calibration,
we can now return to the more traditional and fundamental technique
of observing a single star through a slit-mask which produces 
interference fringes that can be used to determine spatial and
angular calibration independently. 

For all measurements produced in this program we have utilized 
traditional directed vector autocorrelation (hereafter DVA; Bagnuolo
et al.\ 1992) reduction techniques, however, when observing pairs 
with the $``$backup" speckle camera we only observed wider pairs 
where there was no expectation that the observations were 
interferometric. While we called them $``$CCD observations" we had 
an expectation that the results would be better than conventional
CCD astrometry due to the short-exposure time and the many thousand 
correlations obtained in a DVA reduction. Historically, those taken 
with the $``$backup" camera were characterized as $``$CCD 
astrometry" and those taken with the primary camera characterized as
$``$speckle interferometry," even though there were probably speckle
observations which may have been more properly CCD observations and 
vice-versa.

With the new camera capable of observing from the Rayleigh limit of 
the 26$''$ (0\farcs2) to beyond a minute of arc, all reduced with 
the same methodology, a new procedure to characterize the technique 
was needed. Speckle interferometry is most obvious when the measured
separation is less than the diameter of the integrated seeing disk
(e.g., McAlister et al.\ 1987, Figure 2). Observations of pairs with
separations less than the atmospheric seeing ($r_0$) would certainly
be obtained by speckle interferometry. However, when observing 
separated pairs of similar brightness wider than the seeing disk, 
there can be very similar morphologies in their short-exposure 
magnified image structure indicative of isoplanicity, a 
characteristic of speckle images. Based on this and our evaluation
of images taken over the course of the year those observations 
of systems with separations less than or equal to 3\farcs5 are 
characterized as $``$speckle interferometry" while those greater
than 3\farcs5 are $``$CCD astrometry." While there is not a clear 
divider between these two, pairs with separations less than 3\farcs5 
usually demonstrate isoplanicity while those more than 3\farcs5 
usually do not. When the seeing monitor described in Mason et al.\ 
(2017) \S 2.3 comes online we can make a more rigorous 
characterization of observing conditions rather than this subjective 
divider. The pairs which we can now observe include those with shorter 
periods allowing us to improve the orbits of several systems, as 
described in \S 3.3 below.

\section{Results}

With the capability to observe closer pairs our observing list
construction methodology has changed. The initial input list remains
the Washington Double Star (hereafter, WDS, Mason et al.\ 2001) 
Catalog and the main focus of the observing list remains 
neglected pairs, that is, pairs not observed recently or 
unconfirmed. Pairs that are closer in separation tend to be both 
more difficult to observe and move faster, necessitating more 
frequent re-observation. Provided in Table 1 are our observing
cadences and magnitude limits for different separation regimes. In
that table, Column 1 provides the separation range, Column 2 the 
M$_{\rm v}$ limit (of both components), Column 3 the $\Delta$m limit
and Column 4 the time since the last observation for the pair to be 
considered neglected. 

The separation range is based on the last measured value. It is 
likely that we could measure pairs wider than 60$''$, especially 
if their position angle is along a diagonal in our field of view, 
but as these would likely be more slow moving not observing them 
as often is less impactful. While we can observe to the Rayleigh 
limit of 0\farcs2, in evaluating data for the orbit catalog 
(Hartkopf et al.\ 2001) we have found that those obtained close to 
the resolution limit of the telescope/filter combination are less 
valuable for high precision work. Consequently, we set a lower limit
50\% higher at 0\farcs3. The magnitude and magnitude difference is 
based on the WDS values. While we strive to keep these as accurate 
as possible these could be incorrect. If they are, the star(s) tend 
to be fainter than advertised, possibly too faint to observe. These 
are removed from the observing list and their magnitudes updated so 
they should not appear on observing lists in subsequent years. Based
on our experience in 2016 the magnitude limits seem appropriate 
while the magnitude difference limits for those under four 
arcseconds seem too conservative. These will be adjusted to 1.5 and 
1.0, respectively, for the next observing list.

To this list of pairs which are temporally neglected ($N=4390$)
and unconfirmed ($N=59$) additional targets are added based on other
considerations: bright, and thus, potential objects for 
star-trackers ($N=254$), having questionable observations in past 
years ($N=4$), being examined for co-planarity ($N=5$), fast-moving 
($N=19$), systems with multiple determinations in the Sixth Orbit 
Catalog (Hartkopf et al.\ 2001, $N=7$) or those under special 
analysis ($N=18$).

On a given night a pair may be observed multiple times in different
data collection modes and with different magnification as it is not
always obvious which will produce the best result. Further, as 
object acquisition is the longest item in the duty cycle, adding 
additional observations is less consequential. For those 
intranightly observations ($n~=~1257$) the rms values are quite low:
$d\theta~=~0\fdg08$ and $\frac{d\rho}{\rho}~=~0.0017$. A smaller 
number ($n~=~146$) comprise those objects which appear to be slow 
moving\footnote{We assume $\Delta\theta~=~\Delta\rho~=~0$ for 
these.} and were observed on multiple nights. For those internightly 
observations the rms values are twice the intranightly values: 
$d\theta~=~0\fdg15$ and $\frac{d\rho}{\rho}~=~0.0034$. We take these
values as representative of the true error.

\subsection{New Pairs}

Table 2 presents coordinates and magnitude information from 
CDS\footnote{magnitude information is from one of the catalogs 
queried in the Aladin sky atlas, operated at CDS, Strasbourg, 
France. See {\tt http://aladin.u-strasbg.fr/aladin.gml}.} for seven 
pairs which are presented here for the first time. These are closer 
components to known systems or pairs in the same field of view. 
Column one gives the coordinates of the primary of the pair. Column 
two is the WDS identifier while Column three is the discoverer 
designation (where WSI = Washington Stellar Interferometer) number. 
Most of these are additional components to already known pairs. For 
these, we retain the discovery designation of the known pair. 
Columns four and five give the visual magnitudes of the primary and 
secondary, and Column six notes the circumstance of the discovery. 
The mean double star positions of our 26$''$ measures (T, $\theta$, 
and $\rho$) of these systems are given in Table 4. 

As many of these are quite wide we are able to provide between two
and seven additional measures of relative astrometry, in Table 3, 
from other catalogs using the same methodology as described in 
Wycoff et al.\ (2006) and Hartkopf et al.\ (2013). In that table, 
the first two columns identify the system by providing its 
epoch-2000 coordinates and discovery designation (as given in Table 
1). Columns three through five give the epoch of observation 
(expressed as a fractional Julian year), the position angle (in 
degrees), and the separation (in seconds of arc). Note that in all
tables the position angle, measured from North through East, has not
been corrected for precession, and is thus based on the equinox for 
the epoch of observation. Column six and seven is the source of the 
measure and either a reference or note to the source.

\subsection{Measures of Known Pairs}

Tables 4 and 5 present the relative measurements of double stars 
made with the 26$''$ telescope. Table 4 presents those with no 
calculation for motion, either orbital or linear. As in Table 1, the
first two columns identify the system by providing its epoch-2000 
coordinates and discovery designation. Columns three and four give 
the epoch of observation (expressed as a fractional Julian year) and
the position angle (in degrees). Column five gives the position 
angle error. This is the internightly rms value if one is available 
or the mean value of 0\fdg2 if it is not. Columns six and seven 
provide the separation (in seconds of arc) and its error. As above, 
the error is its internightly value or the mean error 
($~=~0.0034\rho$). Column eight is the number of nights in the mean 
position. When this is $``$1" the errors in Columns five and seven 
are the mean results as described above. Finally, Column nine is 
reserved for notes. 

The 2324 measures presented in Table 4 have a mean separation of 
13\farcs897 and a median value of 9\farcs055. The mean number of 
years since the pair was last observed is 7.15. The seven pairs 
listed in Table 2 are included here, as are two objects confirmed 
here: 21473$+$4644 = SRW\phn\phn11AC, a pair first noted by Roger 
Sinnot (1999) and 22384$+$5223 = BAR\phn\phn61BC, a pair first 
measured by E.E.\ Barnard and whose only measure is in the catalog 
of Aitken \& Doolittle (1932), not measured since the discovery 101 
years ago!

Table 5 presents measurements of doubles where some calculation of
position (orbital or linear) is available. The first eight columns 
are the same as Table 4 above. Columns nine \& ten provide the 
O$-$C residual to the determination referenced in Column eleven. 
The final column, like that of Table 4, provides notes. In some 
cases a measure has residuals to more than one calculation. In some 
of those cases the second calculation refers to a new orbit (Table 
6) which is described below.

Not surprisingly, the objects in Table 5 are both closer and more
frequently observed than those of Table 4. The 168 measures 
presented in Table 5 have a mean separation of 12\farcs777 and a 
median value of 5\farcs428. The mean number of years since the pair 
was last observed is 2.59. 

In both Table 4 \& 5 we include older measures from 2002 ($N~=~9$), 
2004 ($12$) and 2005 ($17$) which were inappropriate to include in 
earlier entries in this measurement series. Two pairs, 05013$+$5015 
= STF\phn619 and 21200$+$5259 = STF2789AB, each have multiple orbits
in Kiselev et al.\ (2009). In those cases we are able to identify 
definitively which is the better solution. In addition to those two 
pairs systems are identified in the notes column whose orbital or 
linear solution may need to be improved in the future but where the 
quantity of data is insufficient to improve them at this time.

\subsection{Improved Orbits}

Eight systems with sufficient data to improve their orbits are 
presented in Tables 6 \& 7 and Figures 1 \& 2. All of the individual
measures were weighted by the procedures of Hartkopf et al.\ (2001) 
and calculated with the venerable $``$grid-search" method of 
Hartkopf et al.\ (1989).

Table 6 is broken into two groups. The first orbit we characterize 
as $``$improved but still provisional" and is given without errors. 
They fit the data better than the earlier orbit and should give 
reasonable ephemerides over the next several decades, but the 
elements will all require correction over the course of a complete 
orbit before they can be considered even approximately correct. As 
in earlier tables, the first two columns identify the system by 
providing its epoch-2000 coordinates and discovery designation. 
Columns three through nine provides the seven Campbell elements: 
the period (P in years), the semimajor axis (a$''$ in arcseconds), 
the inclination (i) and longitude of the node ($\Omega$), both in 
degrees, the epoch of the most recent periastron passage (T$_o$ in 
years), the eccentricity (e) and the longitude of periastron 
($\omega$ in degrees). Column ten gives the reference to the 
previous $``$best" orbit and Column eleven the orbital $``$grade" 
following the procedures of Hartkopf et al.\ (2001). 

In the second part of Table 6 are the three orbits we characterize 
as $``$reliable", all with much shorter periods than those in the
first group. All eleven columns are the same as the first part of 
the table, however, here under each element is its formal error. The
precision of the element is defined by the precision of its error. 
Relative visual orbits of all eight systems are plotted in Figure 1,
with the x and y axes indicating the scale in arcseconds. Each solid
curve represents the newly determined orbital elements presented in 
Table 6 and the dashed curve is the orbit of the earlier orbit 
referenced in Column ten. 

\begin{figure}[!ht]
\begin{center}
{\epsfxsize 2.95in \epsffile{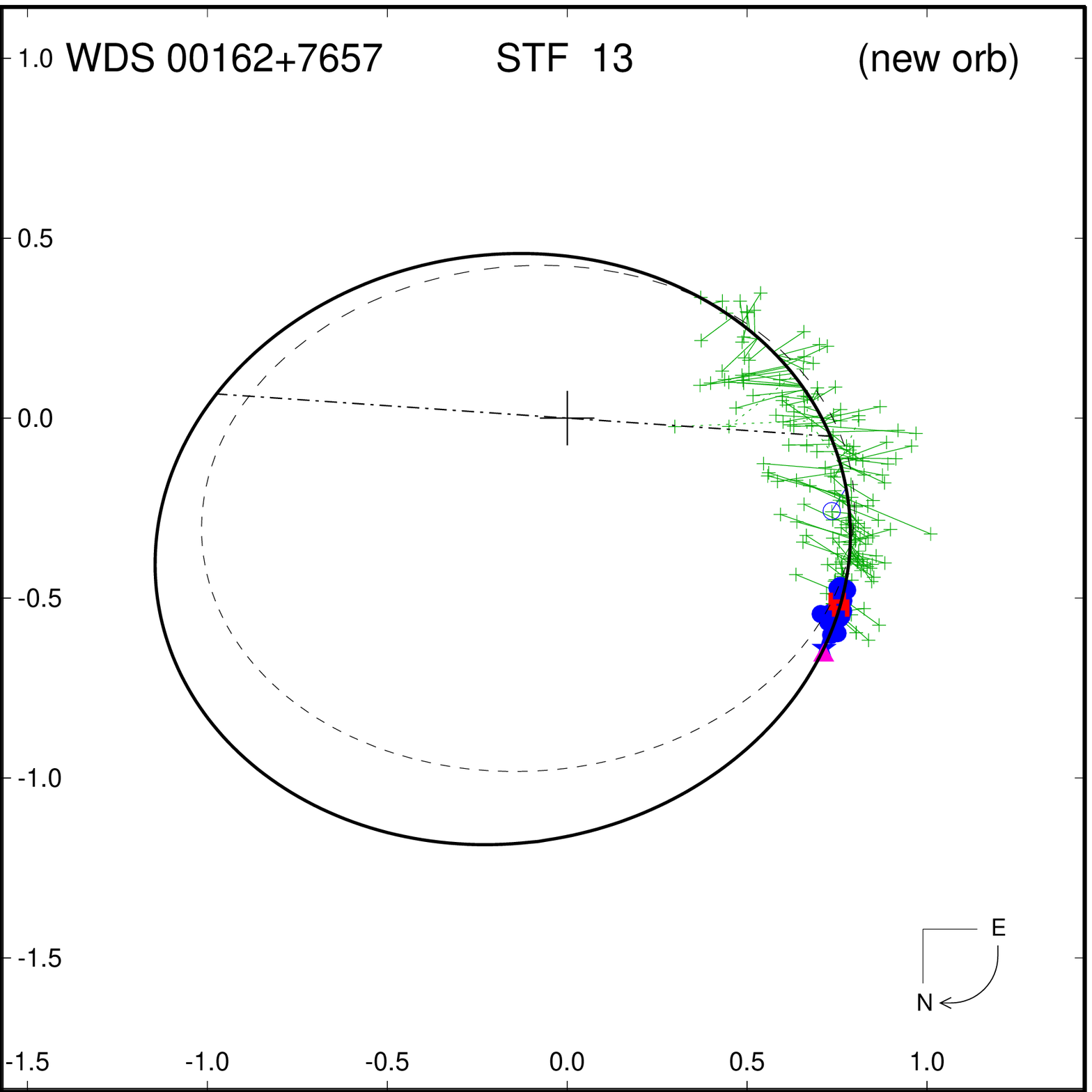} \epsfxsize 2.95in \epsffile{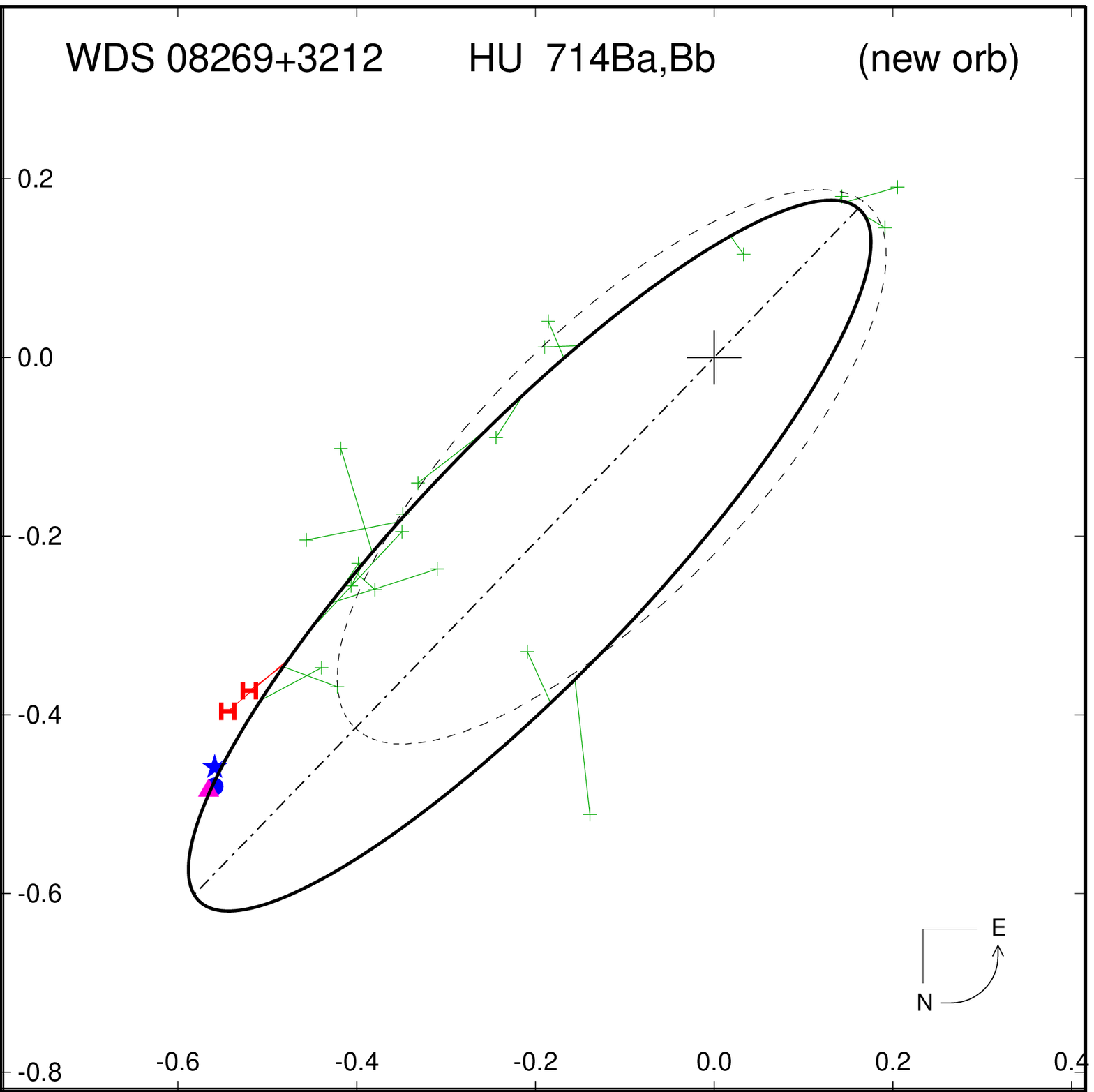}}
{\epsfxsize 2.95in \epsffile{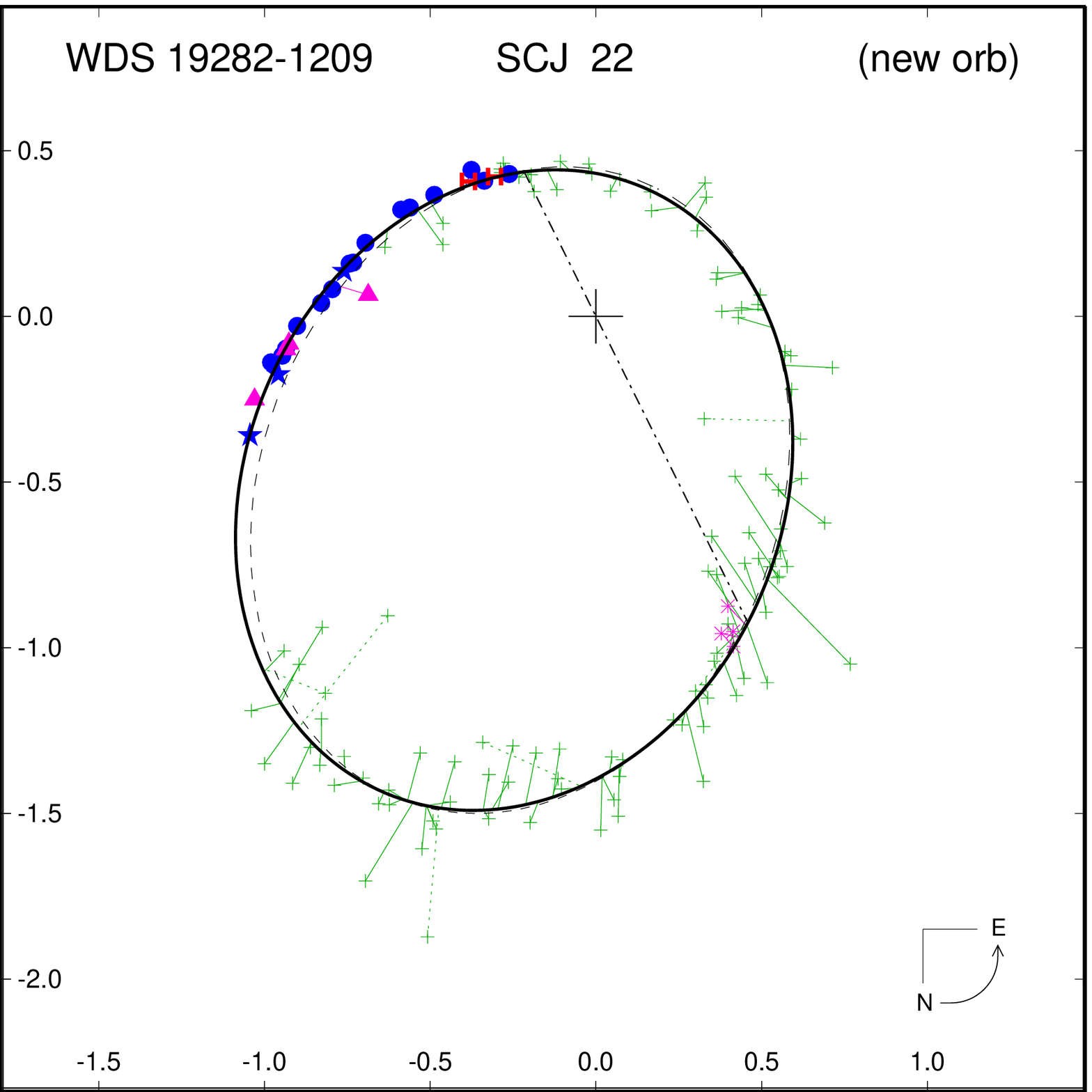} \epsfxsize 2.95in \epsffile{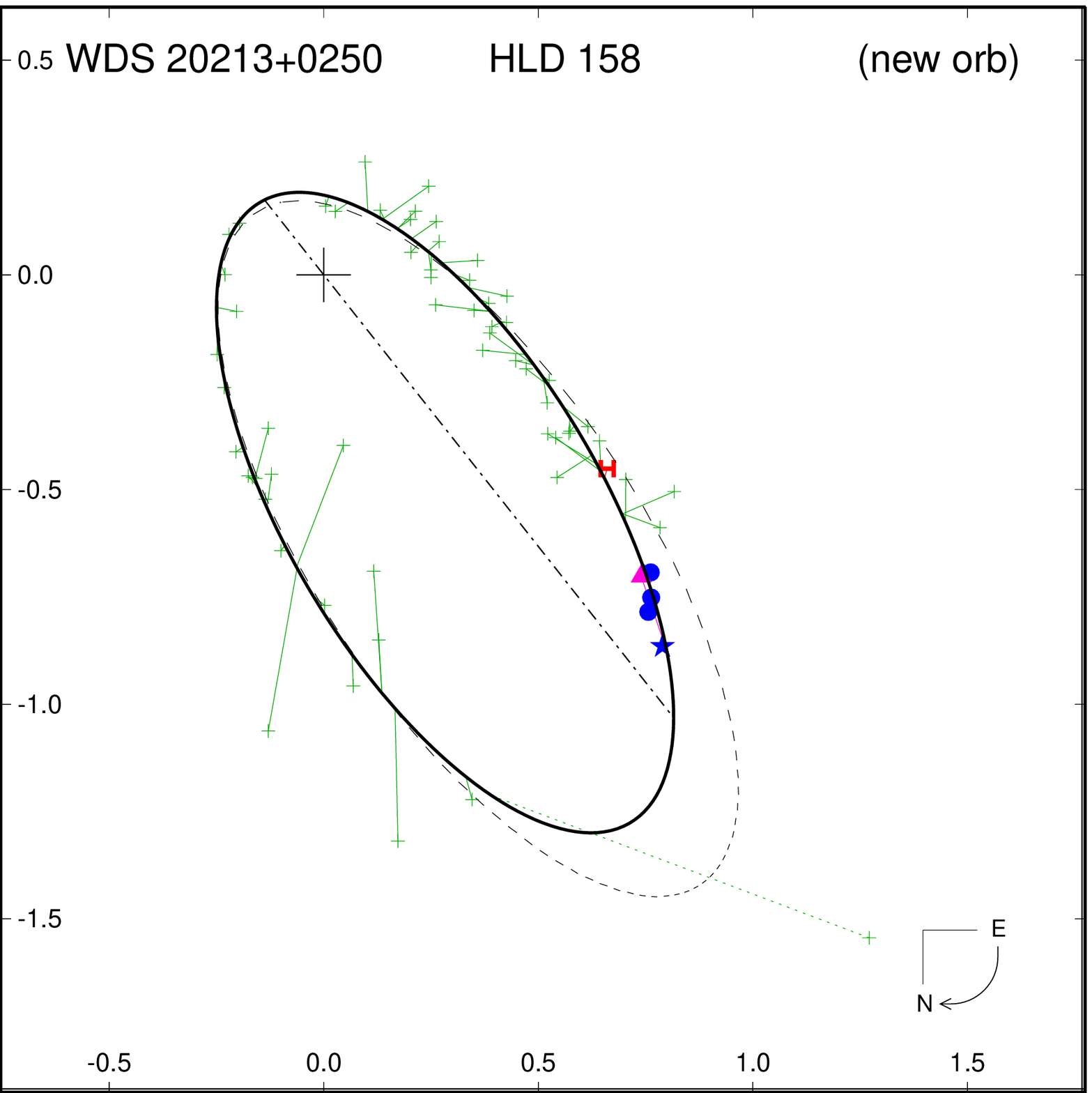}}
\end{center}
\caption{\small Figure 1 illustrates the new orbital solutions, 
plotted together with all published data in the WDS database as well
as the new data in Table 5. In each of these figures, micrometric 
observations are indicated by plus signs, interferometric measures 
by filled circles, space-based measures are indicated by the letter 
`H'.  ``$O-C$" lines connect each measure to its predicted position 
along the new orbit (shown as a thick solid line). Dashed ``$O-C$" 
lines indicate measures given zero weight in the final solution. A 
dot-dash line indicates the line of nodes, and a curved arrow in the
lower right corner of each figure indicates the direction of orbital
motion. The earlier orbit referenced in Table 6 is shown as a dashed
ellipse.}
\end{figure}

Table 7 gives the ephemerides for the pairs in Table 6 over the 
years 2018 through 2026, in two-year increments. Columns 1 and 2 are
the same identifiers as in Table 6, while Columns 3$+$4, 5$+$6, 
etc., through 11$+$12 give predicted values of $\theta$ and $\rho$, 
respectively, for the years 2018.0, 2020.0, etc., through 2026.0. 

Notes to individual systems follow:

{\bf 08269$+$3212 = HU\phn\phn714Ba,Bb} : Also known as 
BD$+$32\phn1731. Based on the period and semi-major axis of Table 6 
and the parallax (17.30$\pm$1.77 mas; van Leeuwen 2007) the mass sum
of this system is 0.775$\pm$0.409 \msun. This is not unreasonable 
for a K2 dwarf and its companion and the orbital elements and mass 
may see significant improvement following the pair reaching its 
widest separation in the apparent orbit ellipse, predicted for 2037.

{\bf 19282$-$1209 = SCJ\phn\phn22} : Also known as HD 183063. Based 
on the period and semi-major axis of Table 6 and the parallax 
(27.31$\pm$0.93 mas; van Leeuwen 2007) the mass sum of this system 
is 1.711$\pm$0.231 \msun, which is the smallest error (15\%) of 
the three here. Completion of a full orbit, which is predicted for 
2044, will allow the period to be very precisely determined in 
following years. 

{\bf 20213$+$0250 = HLD\phn158} : Also known as HD 193651. Based on 
the period and semi-major axis of Table 6 and the parallax 
(2.27$\pm$2.28 mas; van Leeuwen 2007) the worthless mass sum of this
system is 1000$\pm$3138 \msun. The small parallax and its large 
relative error is preventing a more precise mass determination, 
although like 08269$+$3212 above, this one has also yet to reach its
largest separation in the apparent ellipse, predicted for 2067. 

\subsection{Double Stars Not Found}

Table 8 presents four systems which were observed but not detected. 
Possible reasons for nondetection include orbital or differential 
proper motion making the binary too close or too wide to resolve at 
the epoch of observation, a larger than expected $\Delta$m, 
incorrect pointing of the telescope, and misprints and/or errors in 
the original reporting paper. It is hoped that reporting these will 
encourage other double star astronomers to either provide 
corrections to the USNO observations or to verify the lack of 
detection. 

\acknowledgements

This research has also made use of the SIMBAD database, operated at 
CDS, Strasbourg, France and NASA's Astrophysics Data System. The 
continued instrument maintenance by the USNO instrument shop, Gary 
Wieder, Chris Kilian and Phillip Eakens, makes the operation of a 
telescope of this vintage a true delight.


\begin{deluxetable}{cccc}
\tablenum{1}
\tablewidth{0pt}
\tablecaption{Observing List Parameters}
\tablehead{
\colhead{Separation Range} & 
\colhead{M$_{\rm v}$ limit} &
\colhead{$\Delta$M$_{\rm v}$ limit} &
\colhead{Cadence} \nl
}
\startdata
$60\farcs0~>~\rho~\geq~10\farcs0$       & 12    & 3.0 &    10 yr \nl
$10\farcs0~>~\rho~\geq~\phn4\farcs0$    & 11    & 2.0 & \phn5 yr \nl
$\phn4\farcs0~>~\rho~\geq~\phn1\farcs0$ & 10    & 1.0 & \phn2 yr \nl
$\phn1\farcs0~>~\rho~\geq~\phn0\farcs3$ & \phn9 & 0.2 & \phn1 yr \nl
\enddata
\end{deluxetable}


\begin{deluxetable}{lcl@{~}r@{~}lccc}
\tablenum{2}
\tablewidth{0pt}
\tablecaption{New Pairs}
\tablehead{
\colhead{Coordinates} & 
\colhead{WDS} &
\multicolumn{3}{c}{Discoverer} &
\multicolumn{2}{c}{Magnitudes} &
\colhead{Note}  \nl
\colhead{$\alpha,\delta$ (2000)} &
\colhead{Designation} &
\multicolumn{3}{c}{Designation} & 
\colhead{Primary} &
\colhead{Secondary} & 
\colhead{~}
}
\startdata
05\,17\,25.43 $+$03\,53\,13.4 & 05175$+$0354 & BAL & 2142 & AC    & 11.67    &  13.6\phn    & 1   \nl
06\,31\,52.51 $+$04\,56\,15.0 & 06319$+$0457 & SLE &  293 & GH    & 11.75    &  12.77       & 2   \nl
07\,16\,42.20 $+$03\,12\,25.0 & 07167$+$0312 & BAL & 2261 & AC    & 11.4\phn &  13.\phn\phn & 1   \nl
07\,44\,18.79 $-$00\,40\,37.1 & 07443$-$0040 & BAL &  825 & BC    & 11.5\phn &  13.3\phn    & 3   \nl
08\,14\,40.22 $+$04\,38\,21.8 & 08147$+$0438 & WSI &  162 &       & 11.80    &  11.95       &     \nl
20\,41\,13.16 $-$13\,19\,41.8 & 20412$-$1320 & WSI &  163 &       & 10.6\phn &  12.5\phn    & 4   \nl
20\,51\,01.51 $+$39\,15\,05.6 & 20510$+$3915 & SEI & 1279 & BC    & 11.9\phn &  12.5\phn    & 1   \nl
\enddata
\tablenotetext{~}{1 : Physicality status unknown, but closer than the known pair.}
\tablenotetext{~}{2 : AG \& AH both in WDS. This is a closer, smaller $\Delta$m pairing in this
                      multiple system.}
\tablenotetext{~}{3 : This is a common proper motion pair. Based on proper motion the A component
                      appears non-physical.}
\tablenotetext{~}{4 : Magnitudes are red, not visual.}
\end{deluxetable}


\begin{deluxetable}{ll@{~}r@{~}lccclc}
\tablenum{3}
\tablewidth{0pt}
\tablecaption{Data Mining Measures for New Pairs}
\tablehead{
\colhead{WDS or} & 
\multicolumn{3}{c}{Discovery} & 
\colhead{JY} &
\colhead{$\theta$} & 
\colhead{$\rho$} &
\colhead{Source} & 
\colhead{Reference} \nl
\colhead{$\alpha$,$\delta$ (2000)} & 
\multicolumn{3}{c}{Designation} & 
\colhead{~} & 
\colhead{($^{\circ}$)} & 
\colhead{($''$)} &
\colhead{~} &
\colhead{or Note} \nl
}
\startdata
05175$+$0354 & BAL & 2142 & AC & 2000.07\phn\phn    &    289.8 & \phn9.10\phn & 2MASS & 1                      \nl
             &     &      &    & 2000.100\phn       &    289.9 & \phn9.086    & UCAC4 & Zacharias et al.\ 2013 \nl
             &     &      &    & 2010.5\phn\phn\phn &    290.4 & \phn9.36\phn & WISE  & 2                      \nl
06319$+$0457 & SLE &  293 & GH & 1899.13\phn\phn    &    274.6 &    18.704    & AC    & Urban et al.\ 1998     \nl
             &     &      &    & 1932.09\phn\phn    &    274.8 &    18.814    & AC    & Urban et al.\ 1998     \nl
             &     &      &    & 1999.87\phn\phn    &    274.7 &    18.53\phn & 2MASS & 1                      \nl
             &     &      &    & 2000.146\phn       &    274.7 &    18.552    & UCAC4 & Zacharias et al.\ 2013 \nl
             &     &      &    & 2010.5\phn\phn\phn &    274.7 &    18.75\phn & WISE  & 2                      \nl
07167$+$0312 & BAL & 2261 & AC & 1999.95\phn\phn    & \phn49.3 & \phn4.47\phn & 2MASS & 1                      \nl
             &     &      &    & 2000.152\phn       & \phn49.1 & \phn4.437    & UCAC4 & Zacharias et al.\ 2013 \nl
07443$-$0040 & BAL &  825 & BC & 1998.87\phn\phn    &    287.3 &    17.30\phn & 2MASS & 1                      \nl
             &     &      &    & 1999.2087          &    287.1 &    17.30\phn & DENIS & 3                      \nl
             &     &      &    & 1999.9514          &    286.8 &    17.30\phn & DENIS & 3                      \nl
             &     &      &    & 2000.126\phn       &    287.2 &    17.292    & UCAC4 & Zacharias et al.\ 2013 \nl
08147$+$0438 & WSI &  162 &    & 1899.15\phn\phn    &    248.1 &    20.018    & AC    & Urban et al.\ 1998     \nl
             &     &      &    & 1910.20\phn\phn    &    248.4 &    20.155    & AC    & Urban et al.\ 1998     \nl
             &     &      &    & 1918.17\phn\phn    &    248.4 &    20.180    & AC    & Urban et al.\ 1998     \nl
             &     &      &    & 1991.53\phn\phn    &    249.9 &    20.413    & Tycho & H$\o$g et al.\ 2000    \nl
             &     &      &    & 2000.07\phn\phn    &    249.4 &    20.54\phn & 2MASS & 1                      \nl
             &     &      &    & 2000.183\phn       &    249.5 &    20.594    & UCAC4 & Zacharias et al.\ 2013 \nl
             &     &      &    & 2010.5\phn\phn\phn &    249.5 &    20.59\phn & WISE  & 2                      \nl
20412$-$1320 & WSI &  163 &    & 1996.6929          & \phn47.1 &    10.31\phn & DENIS & 3                      \nl
             &     &      &    & 2000.37\phn\phn    & \phn46.8 &    10.259    & UCAC4 & Zacharias et al.\ 2013 \nl
             &     &      &    & 2000.67\phn\phn    & \phn46.6 &    10.26\phn & 2MASS & 1                      \nl
20510$+$3915 & SEI & 1279 & BC & 1998.46\phn\phn    &    191.0 & \phn7.72\phn & 2MASS & 1                      \nl
             &     &      &    & 2002.671\phn       &    190.8 & \phn7.690    & UCAC4 & Zacharias et al.\ 2013 \nl
\enddata
\tablenotetext{~}{1 : Cutri et al.\ (2003), All-sky Release. See {\tt Vizier On-line Data Catalog: II/246}.}
\tablenotetext{~}{2 : Cutri et al.\ (2012), All-Sky Release. See {\tt Vizier On-line Data Catalog: II/311}.}
\tablenotetext{~}{3 : DENIS Consortium (2005), Third data release (20 Sep 2005). See {\tt Vizier On-line Data Catalog: II/263}.}
\end{deluxetable}

\input{tab4}


\begin{deluxetable}{ll@{~}r@{~}lcrrrrccclc}
\tiny
\tablenum{5}
\tablewidth{0pt}
\tablecaption{ICCD Measurements of Double Stars with Orbit \& Linear Residuals}
\tablehead{
\colhead{WDS Desig.} & 
\multicolumn{3}{c}{Discoverer} &
\colhead{JY} &
\colhead{$\theta$} & 
\colhead{$\sigma\theta$} & 
\colhead{$\rho$} &
\colhead{$\sigma\rho$} &
\colhead{n} &
\colhead{O$-$C$_{\theta}$} &
\colhead{O$-$C$_{\rho}$} &
\colhead{Reference} &
\colhead{Note} \nl
\colhead{$\alpha,\delta$ (2000)} &
\multicolumn{3}{c}{Designation} & 
\colhead{~} & 
\colhead{($\circ$)} & 
\colhead{($\circ$)} & 
\colhead{($''$)} &
\colhead{($''$)} &
\colhead{~} &
\colhead{($\circ$)} & 
\colhead{($''$)} &
\colhead{~} &
\colhead{~} \nl
}
\startdata
00014$+$3937 & HLD &   60 &       & 2016.850 & 166.6 & 0.2 &  1.340 & 0.005 & 1 & \phs1.1 & \phs0.032 & Hartkopf \& Mason 2011a     & A \nl
00032$+$4508 & HJ  & 1927 &       & 2016.853 &  73.1 & 0.2 &  9.590 & 0.033 & 1 & \phs0.4 &  $-$0.028 & Hartkopf \& Mason 2011b     &   \nl
00057$+$4549 & STT &  547 & AB    & 2016.853 & 189.2 & 0.2 &  5.984 & 0.020 & 1 &  $-$0.2 & \phs0.084 & Popovic \& Pavlovic 1996    & A \nl
             &     &      &       &          &       &     &        &       &   & \phs0.3 &  $-$0.006 & Kiyaeva et al.\ 2001        &   \nl
00162$+$7657 & STF &   13 &       & 2016.785 &  48.2 & 0.2 &  0.958 & 0.003 & 1 & \phs0.3 & \phs0.050 & 0levic \& Jovanovic 2001    &   \nl
             &     &      &       &          &       &     &        &       &   &  $-$0.4 & \phs0.002 & Table 6                     &   \nl
00209$+$3259 & AC  &    1 &       & 2016.850 & 289.9 & 0.2 &  1.846 & 0.006 & 1 & \phs1.0 & \phs0.004 & Zirm 2015                   &   \nl
00272$+$4959 & STF &   30 & AB    & 2016.850 & 315.8 & 0.2 & 13.249 & 0.045 & 1 & \phs0.9 &  $-$0.054 & Hartkopf \& Mason 2011b     &   \nl
00287$+$3718 & A   & 1504 & AB    & 2016.850 &  49.3 & 0.2 &  0.581 & 0.002 & 1 & \phs4.0 &  $-$0.015 & Zirm 2014                   & A \nl
00360$+$2959 & STF &   42 & AB    & 2016.815 &  20.8 & 0.2 &  6.265 & 0.021 & 1 & \phs0.5 &  $-$0.050 & Kiselev et al.\ 2009        &   \nl
00360$+$2959 & STF &   42 & AC    & 2016.815 & 289.6 & 0.2 & 36.681 & 0.125 & 1 & \phs0.6 & \phs0.039 & Hartkopf \& Mason 2011b     &   \nl
00384$+$4059 & STF &   44 &       & 2016.779 & 274.2 & 0.2 & 12.967 & 0.044 & 1 & \phs0.2 & \phs0.199 & Hartkopf \& Mason 2011b     & B \nl
00464$+$3057 & STFA&    1 & AB    & 2016.815 &  46.5 & 0.2 & 47.569 & 0.162 & 1 & \phs0.3 & \phs0.244 & Hartkopf \& Mason 2011b     &   \nl
00521$+$1036 & STF &   67 &       & 2016.839 & 349.2 & 0.2 &  2.313 & 0.008 & 1 &  $-$0.1 & \phs0.026 & Hartkopf \& Mason 2011a     & A \nl
00594$+$0047 & STF &   80 & AB    & 2016.849 & 338.9 & 0.2 & 29.801 & 0.025 & 2 & \phs0.3 & \phs0.145 & Hartkopf \& Mason 2011b     &   \nl
01032$+$2006 & LDS &  873 &       & 2004.852 &  51.2 & 0.2 &  2.556 & 0.009 & 2 & \phs0.1 &  $-$0.033 & Rica et al.\ 2012           & A,C \nl
01048$-$0528 & STF &   86 & AB    & 2016.776 & 137.3 & 0.3 & 16.847 & 0.014 & 2 & \phs0.2 & \phs0.061 & Hartkopf \& Mason 2011b     &   \nl
01198$-$0031 & STF &  113 & A,BC  & 2016.849 &  21.4 & 0.2 &  1.623 & 0.008 & 2 & \phs0.3 &  $-$0.015 & Zirm 2015                   &   \nl
01200$-$1549 & HJ  & 2036 &       & 2016.818 & 338.1 & 0.2 &  2.398 & 0.008 & 1 & \phs0.4 & \phs0.028 & Olevic et al.\ 2003         & A \nl
01207$+$4620 & STF &  112 & AB    & 2016.850 & 337.4 & 0.2 & 19.042 & 0.065 & 1 & \phs0.8 & \phs0.101 & Hartkopf \& Mason 2011b     &   \nl
01211$+$6439 & S   &  397 &       & 2016.850 & 342.1 & 0.2 & 57.396 & 0.195 & 1 & \phs0.7 & \phs0.098 & Hartkopf \& Mason 2011b     &   \nl
01399$+$1515 & STF &  142 & AB    & 2016.818 &  68.1 & 0.2 & 23.311 & 0.079 & 1 & \phs0.2 & \phs0.249 & Hartkopf \& Mason 2011b     & B \nl
01456$-$2503 & HJ  & 3461 & AB    & 2002.900 &  23.1 & 0.2 &  4.898 & 0.017 & 1 &  $-$1.5 &  $-$0.050 & Mason \& Hartkopf 2014      & A,C \nl
01467$+$3310 & STF &  158 & AB    & 2016.883 & 271.5 & 0.2 &  2.230 & 0.008 & 1 &  $-$1.5 & \phs0.213 & Hartkopf \& Mason 2011a     & A \nl
01488$-$0125 & STF &  171 & AB    & 2016.891 & 165.4 & 0.2 & 34.513 & 0.117 & 1 & \phs0.7 & \phs0.197 & Hartkopf \& Mason 2011b     &   \nl
01493$+$4754 & STF &  162 & AB    & 2016.967 & 199.0 & 0.0 &  1.898 & 0.034 & 2 & \phs0.3 &  $-$0.059 & Zirm \& Rica 2014           & B \nl
             &     &      &       &          &       &     &        &       &   & \phs0.3 &  $-$0.059 & Genet et al.\ 2015          &   \nl
01510$+$2107 & STF &  175 & AB    & 2016.891 &   0.2 & 0.2 & 28.381 & 0.096 & 1 & \phs0.8 & \phs0.056 & Hartkopf \& Mason 2011b     &   \nl
02216$+$2338 & STF &  254 &       & 2016.926 &  16.5 & 0.1 & 12.111 & 0.013 & 2 & \phs0.4 & \phs0.003 & Hartkopf \& Mason 2011b     &   \nl
02231$+$7021 & MLR &  377 & AB    & 2004.885 & 144.4 & 0.2 &  0.655 & 0.002 & 2 &  $-$1.5 & \phs0.084 & Pavlovic \& Todorovic 2005  & C \nl
02407$+$6117 & STF &  284 & AB    & 2016.836 & 191.2 & 0.3 &  6.906 & 0.013 & 2 & \phs1.0 &  $-$0.023 & Hartkopf \& Mason 2011b     &   \nl
02475$+$1922 & STF &  305 & AB    & 2016.856 & 307.6 & 0.0 &  3.645 & 0.004 & 3 & \phs0.6 & \phs0.043 & Mason \& Hartkopf 2014      & A \nl
02558$+$3429 & STF &  325 & AB    & 2016.846 & 147.4 & 0.0 & 23.261 & 0.014 & 2 & \phs1.1 & \phs0.094 & Hartkopf \& Mason 2011b     &   \nl
02563$+$7253 & STF &  312 & AB    & 2016.875 &  48.8 & 0.2 &  1.777 & 0.006 & 1 & \phs2.5 & \phs0.017 & Cvetkovic \& Novakovic 2006 &   \nl
02592$+$2120 & STF &  333 & AB    & 2016.888 & 210.6 & 0.0 &  1.371 & 0.002 & 3 & \phs0.8 & \phs0.038 & Rica 2012                   & A \nl
03121$-$2859 & HJ  & 3555 &       & 2002.900 & 299.1 & 0.2 &  4.880 & 0.017 & 1 &  $-$0.1 &  $-$0.127 & S\"{o}derhjelm 1999         & A,C \nl
03127$+$7133 & STT &   50 & AB    & 2016.875 & 146.2 & 0.2 &  1.026 & 0.003 & 1 & \phs3.0 & \phs0.132 & Scardia et al.\ 2012        & A \nl
03140$+$0044 & STF &  367 &       & 2016.935 & 129.8 & 0.2 &  1.232 & 0.004 & 1 &  $-$0.2 &  $-$0.013 & Riddle et al.\ 2015         & A \nl
03196$+$6714 & HU  & 1056 &       & 2016.881 &  80.6 & 0.2 &  1.035 & 0.004 & 1 & \phs0.8 &  $-$0.027 & Zirm 2015                   & A \nl
03207$+$4641 & BU  & 1294 &       & 2016.990 & 240.1 & 0.2 &  8.714 & 0.030 & 1 & \phs1.1 & \phs0.033 & Hartkopf \& Mason 2011b     &   \nl
03217$+$0845 & STF &  380 &       & 2016.935 &   6.6 & 0.2 &  0.923 & 0.003 & 1 & \phs4.2 & \phs0.041 & Popovic \& Pavlovic 1996    & A \nl
03480$+$6840 & WNO &   16 & AD    & 2016.051 & 352.4 & 0.2 & 30.446 & 0.104 & 1 & \phs1.1 & \phs0.093 & Hartkopf \& Mason 2011b     &   \nl
04422$+$3731 & STF &  577 &       & 2016.990 & 331.8 & 0.2 &  0.620 & 0.002 & 1 & \phs4.2 &  $-$0.062 & Riddle et al.\ 2015         & A \nl
04465$+$7128 & HJ  & 2235 &       & 2016.853 & 156.0 & 0.2 & 40.587 & 0.138 & 1 & \phs0.9 & \phs0.204 & Hartkopf \& Mason 2011b     &   \nl
05013$+$5015 & STF &  619 &       & 2016.037 & 161.2 & 0.2 &  4.126 & 0.014 & 1 & \phs1.9 & \phs0.034 & Kiselev et al.\ 2009        & D \nl
05047$+$7404 & STT &   89 &       & 2002.996 & 304.0 & 0.2 &  0.329 & 0.001 & 1 & \phs0.6 &  $-$0.018 & Alzner 1998                 & A,C \nl
05364$+$2200 & STF &  742 &       & 2016.990 & 275.4 & 0.2 &  4.152 & 0.014 & 1 &  $-$0.3 & \phs0.003 & Hopmann 1973                &   \nl
05446$+$6320 & STI &  579 &       & 2016.051 & 121.2 & 0.2 &  8.389 & 0.029 & 1 &  $-$1.6 & \phs0.455 & Hurowitz et al.\ 2013       & B \nl
06482$+$5542 & STF &  958 & AB    & 2016.212 &  76.2 & 0.2 &  4.159 & 0.014 & 1 &  $-$0.7 &  $-$0.367 & Kiselev et al.\ 2009        &   \nl
             &     &      &       &          &       &     &        &       &   &  $-$0.3 &  $-$0.297 & Kiselev et al.\ 2009        &   \nl
06487$+$0737 & A   & 2731 & AB    & 2005.170 &  64.4 & 0.2 &  1.195 & 0.004 & 1 & \phs0.3 &  $-$0.088 & Prieur et al.\ 2012         & C \nl
07201$+$2159 & STF & 1066 &       & 2002.172 & 224.1 & 0.2 &  5.889 & 0.020 & 2 &  $-$1.4 & \phs0.160 & Hopmann 1960                & A,C \nl
08122$+$1739 & STF & 1196 & AB,C  & 2004.906 &  72.6 & 0.2 &  6.013 & 0.020 & 2 & \phs1.9 & \phs0.105 & Heintz 1996                 & A,C \nl
08122$+$1739 & STF & 1196 & AB,C  & 2005.287 &  71.8 & 0.2 &  6.262 & 0.021 & 1 & \phs1.4 & \phs0.354 & Heintz 1996                 & C \nl
08122$+$1739 & STF & 1196 & AB,C  & 2016.832 &  64.2 & 0.2 &  5.959 & 0.020 & 1 &  $-$1.2 & \phs0.030 & Heintz 1996                 &   \nl
08269$+$3212 & HU  &  714 & Ba,Bb & 2004.160 & 309.4 & 0.2 &  0.723 & 0.002 & 1 &  $-$6.3 & \phs0.146 & Baize 1993                  & C \nl
             &     &      &       &          &       &     &        &       &   &  $-$0.0 & \phs0.003 & Table 6                     &   \nl
08369$+$2315 & AG  &  154 &       & 2016.832 &   1.3 & 0.2 &  2.678 & 0.009 & 1 & \phs0.4 & \phs0.002 & Hartkopf \& Mason 2011c     &   \nl
08568$-$1726 & ARG &   72 & AB    & 2016.996 & 182.6 & 0.2 &  4.053 & 0.014 & 1 & \phs0.6 & \phs0.029 & Hartkopf \& Mason 2011b     &   \nl
09013$+$1516 & STF & 1300 & AB    & 2016.163 & 177.5 & 0.2 &  5.029 & 0.017 & 1 &  $-$1.4 &  $-$0.005 & Zirm 2008                   &   \nl
09079$-$0708 & STF & 1316 & AC    & 2016.172 & 279.5 & 0.3 &  7.949 & 0.010 & 2 &  $-$0.1 & \phs0.031 & Hartkopf \& Mason 2011b     &   \nl
09079$-$0708 & STF & 1316 & BC    & 2016.204 & 297.1 & 0.2 & 14.218 & 0.048 & 1 & \phs2.2 & \phs0.104 & Friedman et al.\ 2012       &   \nl
09144$+$5241 & STF & 1321 & AB    & 2016.213 &  98.1 & 0.2 & 17.102 & 0.058 & 1 &  $-$0.1 & \phs0.255 & Chang 1972                  & A \nl
09157$-$0114 & STF & 1329 & AB    & 2016.196 & 265.4 & 0.1 &  8.301 & 0.036 & 2 &  $-$0.0 & \phs0.046 & Hartkopf \& Mason 2011b     &   \nl
10160$+$1200 & HJ  &  156 & BC    & 2016.163 & 355.3 & 0.2 & 29.664 & 0.101 & 1 & \phs0.2 &  $-$0.162 & Hurowitz et al.\ 2014       &   \nl
10200$+$1950 & STF & 1424 & AB    & 2016.213 & 125.8 & 0.2 &  4.898 & 0.017 & 1 &  $-$1.4 & \phs0.163 & Romanenko \& Kiselev 2014   & A \nl
             &     &      &       &          &       &     &        &       &   &  $-$0.6 & \phs0.183 & Romanenko \& Kiselev 2014   &   \nl
11268$+$0301 & STF & 1540 & AB    & 2016.163 & 149.6 & 0.2 & 28.086 & 0.095 & 1 & \phs2.8 &  $-$0.532 & Hopmann 1960                &   \nl
11279$+$0251 & STFA&   19 & AB    & 2016.169 & 181.6 & 0.2 & 88.411 & 0.301 & 1 &  $-$0.2 &  $-$0.339 & Hartkopf \& Mason 2011b     &   \nl
12078$+$3110 & SEI &  527 &       & 2016.185 &   6.7 & 0.2 & 42.253 & 0.144 & 1 & \phs0.2 &  $-$0.098 & Hartkopf \& Mason 2011b     &   \nl
13379$+$4808 & ES  &  608 & AB    & 2002.383 & 323.5 & 0.2 &  1.942 & 0.007 & 2 & \phs3.8 & \phs0.009 & Seymour et al.\ 2002        & A,C \nl
13379$+$4808 & ES  &  608 & AB    & 2005.403 & 322.1 & 0.2 &  1.752 & 0.006 & 1 &  $-$0.2 &  $-$0.115 & Seymour et al.\ 2002        & C \nl
13550$-$0804 & STF & 1788 & AB    & 2005.356 &  98.7 & 0.2 &  3.635 & 0.012 & 1 & \phs0.1 & \phs0.104 & Hopmann 1970                & A,C \nl
14203$+$4830 & STF & 1834 &       & 2016.412 & 105.4 & 0.2 &  1.656 & 0.006 & 1 & \phs3.0 & \phs0.050 & Hartkopf \& Mason 2015      & A \nl
14410$+$5757 & STF & 1872 & AB    & 2016.467 &  50.8 & 0.2 &  7.602 & 0.026 & 1 & \phs0.6 & \phs0.033 & Kiyaeva et al.\ 2010        &   \nl
14463$+$0939 & STF & 1879 & AB    & 2016.412 &  82.6 & 0.2 &  1.720 & 0.006 & 1 &  $-$0.2 & \phs0.000 & Mason et al.\ 1999          &   \nl
14464$-$0723 & STF & 1876 & AB    & 2016.412 & 114.9 & 0.2 &  1.237 & 0.004 & 1 & \phs1.5 &  $-$0.006 & Seymour et al.\ 2002        &   \nl
14497$+$4843 & STF & 1890 &       & 2016.467 &  46.8 & 0.2 &  2.659 & 0.009 & 1 & \phs1.0 & \phs0.077 & Hartkopf \& Mason 2011b     & B \nl
14534$+$1542 & STT &  288 &       & 2016.412 & 160.5 & 0.2 &  1.006 & 0.003 & 1 & \phs2.8 & \phs0.013 & Heintz 1998                 & A \nl
15038$+$4739 & STF & 1909 &       & 2016.467 &  71.4 & 0.2 &  0.762 & 0.003 & 1 &  $-$0.2 &  $-$0.003 & Zirm 2011                   &   \nl
15382$+$3615 & STF & 1964 & CD    & 2016.451 &  21.8 & 0.2 &  1.546 & 0.005 & 1 & \phs2.0 & \phs0.038 & Drummond et al.\ 1995       &   \nl
16060$+$1319 & STF & 2007 & AB    & 2016.544 & 323.7 & 0.2 & 38.992 & 0.133 & 1 & \phs1.8 & \phs0.615 & Hartkopf \& Mason 2011b     & B,C \nl
16081$+$1703 & STF & 2010 & AB    & 2016.544 &  15.4 & 0.2 & 27.280 & 0.093 & 1 & \phs1.8 & \phs0.244 & Hartkopf \& Mason 2011b     &   \nl
16147$+$3352 & STF & 2032 & AB    & 2016.544 & 240.8 & 0.2 &  7.369 & 0.025 & 1 & \phs2.3 & \phs0.161 & Raghavan et al.\ 2009       & A \nl
16160$+$0721 & STF & 2026 & AB    & 2016.544 &  18.9 & 0.2 &  3.573 & 0.012 & 1 & \phs2.0 & \phs0.049 & Scardia et al.\ 2011        & A \nl
16289$+$1825 & STF & 2052 & AB    & 2016.544 & 121.0 & 0.2 &  2.422 & 0.008 & 1 & \phs2.2 & \phs0.024 & Scardia et al.\ 2015c       &   \nl
16439$+$4329 & D   &   15 &       & 2002.467 &  92.9 & 0.2 &  0.425 & 0.001 & 4 &  $-$3.5 &  $-$0.014 & Alzner 2007                 & A,C \nl
17010$+$6807 & MLR &  199 & AB    & 2016.481 &  76.5 & 0.1 & 19.245 & 0.014 & 2 & \phs1.7 &  $-$0.061 & Hartkopf \& Mason 2011b     &   \nl
17053$+$5428 & STF & 2130 & AB    & 2016.467 &   3.2 & 0.2 &  2.550 & 0.009 & 1 & \phs1.7 & \phs0.026 & Prieur et al.\ 2012         & A \nl
17146$+$1423 & STF & 2140 & AB    & 2005.663 & 104.2 & 0.2 &  4.806 & 0.016 & 4 & \phs0.3 & \phs0.160 & Baize 1978                  & A,C \nl
17146$+$1423 & STF & 2140 & AB    & 2016.642 & 103.5 & 0.2 &  4.832 & 0.016 & 1 & \phs0.7 & \phs0.190 & Baize 1978                  &   \nl
17419$+$7209 & STF & 2241 & AB    & 2016.653 &  16.2 & 0.2 & 30.122 & 0.102 & 1 &  $-$0.8 & \phs0.544 & Kiselev et al.\ 2009        & A \nl
17520$+$1520 & STT &  338 & AB    & 2016.544 & 167.7 & 0.2 &  0.816 & 0.003 & 1 & \phs4.1 &  $-$0.013 & Prieur et al.\ 2012         & A \nl
18002$+$8000 & STF & 2308 & AB    & 2016.653 & 231.8 & 0.2 & 18.736 & 0.064 & 1 &  $-$0.1 &  $-$0.071 & Kiselev \& Romanenko 1996   &   \nl
             &     &      &       &          &       &     &        &       &   &  $-$0.1 & \phs0.006 & Hartkopf \& Mason 2011b     &   \nl
18032$+$2522 & STF & 2268 & AC    & 2016.645 & 201.9 & 0.2 & 24.521 & 0.083 & 1 &  $-$0.5 & \phs0.440 & Hartkopf \& Mason 2011b     & B \nl
18044$+$0329 & STF & 2266 & AC    & 2016.552 & 203.0 & 0.2 & 50.802 & 0.173 & 1 & \phs2.0 & \phs0.088 & Hartkopf \& Mason 2011b     &   \nl
18055$+$0230 & STF & 2272 & AB    & 2016.552 & 127.2 & 0.2 &  6.448 & 0.022 & 1 & \phs2.6 & \phs0.048 & Eggenberger et al.\ 2008    &   \nl
18101$+$1629 & STF & 2289 &       & 2016.552 & 220.6 & 0.2 &  1.215 & 0.004 & 1 & \phs5.0 &  $-$0.026 & Hopmann 1964                & A \nl
18146$+$0011 & STF & 2294 &       & 2016.642 &  93.3 & 0.2 &  1.362 & 0.005 & 1 & \phs0.1 & \phs0.034 & Scardia et al.\ 2015b       & A \nl
18312$+$1311 & STF & 2330 &       & 2016.664 & 165.4 & 0.2 & 16.386 & 0.056 & 1 &  $-$0.1 & \phs0.123 & Hartkopf \& Mason 2011b     &   \nl
18443$+$3940 & STF & 2382 & AB    & 2016.689 & 345.9 & 0.2 &  2.297 & 0.008 & 1 & \phs0.3 &  $-$0.035 & Mason et al.\ 2004          & A \nl
             &     &      &       &          &       &     &        &       &   & \phs0.4 & \phs0.047 & Novakovic \& Todorovic 2006 &   \nl
18443$+$3940 & STF & 2383 & CD    & 2016.689 &  76.3 & 0.2 &  2.404 & 0.008 & 1 & \phs0.8 & \phs0.014 & Docobo \& Costa 1984        &   \nl
18489$+$1615 & STF & 2400 & A,BC  & 2016.648 & 160.8 & 0.2 & 11.034 & 0.038 & 1 &  $-$0.2 & \phs0.229 & Cvetkovic et al.\ 2016      & B \nl
19022$+$0845 & STF & 2436 & AB    & 2016.642 & 314.9 & 0.2 & 30.296 & 0.103 & 1 &  $-$0.1 & \phs0.212 & Hartkopf \& Mason 2011b     &   \nl
19027$-$0043 & STF & 2434 & A,BC  & 2016.642 &  88.8 & 0.2 & 27.726 & 0.094 & 1 &  $-$0.3 & \phs0.112 & Hartkopf \& Mason 2011b     &   \nl
19121$+$4951 & STF & 2486 & AB    & 2005.720 & 205.7 & 0.2 &  7.300 & 0.025 & 1 & \phs0.0 &  $-$0.109 & Hale 1994                   & A,C \nl
19121$+$4951 & STF & 2486 & AB    & 2016.702 & 204.2 & 0.2 &  7.216 & 0.025 & 1 & \phs0.2 & \phs0.004 & Hale 1994                   &   \nl
19169$+$6312 & STF & 2509 &       & 2016.700 & 328.2 & 0.2 &  1.869 & 0.006 & 1 &  $-$0.3 & \phs0.028 & Zirm 2014                   & A \nl
19266$+$2530 & STF & 2524 & AB    & 2016.645 &  81.9 & 0.2 &  5.427 & 0.018 & 1 &  $-$0.4 & \phs0.014 & Hartkopf \& Mason 2011b     &   \nl
19266$+$2719 & STF & 2525 & AB    & 2016.645 & 289.2 & 0.2 &  2.223 & 0.008 & 1 &  $-$0.2 & \phs0.045 & Scardia et al.\ 2015a       & A \nl
19282$-$1209 & SCJ &   22 &       & 2016.653 & 288.9 & 0.2 &  1.104 & 0.004 & 1 &  $-$3.1 & \phs0.016 & Docobo \& Ling 2009         &   \nl
             &     &      &       &          &       &     &        &       &   &  $-$0.1 &  $-$0.002 & Table 6                     &   \nl
19368$-$1027 & STF & 2541 &       & 2016.656 & 326.1 & 0.2 &  5.500 & 0.019 & 1 &  $-$0.3 &  $-$0.197 & Hartkopf \& Mason 2011b     & B \nl
20014$+$1045 & STF & 2613 & AB    & 2016.664 & 355.1 & 0.2 &  3.581 & 0.012 & 1 & \phs3.0 &  $-$0.564 & Hopmann 1973                & A \nl
20099$+$2055 & STF & 2637 & AC    & 2016.814 & 221.5 & 0.2 & 91.482 & 0.311 & 1 & \phs0.2 & \phs0.456 & Hartkopf \& Mason 2011b     &   \nl
20099$+$2055 & STF & 2637 & BC    & 2016.814 & 215.0 & 0.2 & 95.916 & 0.326 & 1 &  $-$0.3 & \phs0.439 & Hartkopf \& Mason 2011b     &   \nl
20129$+$0959 & HJ  &  908 &       & 2016.814 & 350.3 & 0.2 & 32.528 & 0.111 & 1 & \phs0.4 &  $-$0.461 & Hartkopf \& Mason 2011b     & B \nl
20213$+$0250 & HLD &  158 &       & 2016.875 &  42.3 & 0.2 &  1.171 & 0.004 & 1 &  $-$4.1 &  $-$0.068 & Heintz 1998                 &   \nl
             &     &      &       &          &       &     &        &       &   & \phs0.3 &  $-$0.026 & Table 6                     &   \nl
20312$+$0513 & AG  &  257 &       & 2016.645 &  74.9 & 0.2 &  1.659 & 0.006 & 1 & \phs0.3 & \phs0.046 & Zasche et al.\ 2009         & A \nl
20352$+$0608 & HWE &   98 &       & 2016.645 &   2.1 & 0.2 & 58.516 & 0.199 & 1 &  $-$0.3 & \phs0.397 & Hartkopf \& Mason 2011b     &   \nl
20387$+$3838 & STF & 2708 & AB    & 2016.853 & 323.3 & 0.2 & 57.444 & 0.195 & 1 & \phs0.5 & \phs0.131 & Hartkopf \& Mason 2011b     &   \nl
20450$+$1244 & BU  &   64 & AB    & 2002.602 & 350.0 & 0.2 &  0.678 & 0.002 & 3 &  $-$1.0 & \phs0.018 & Brendley \& Hartkopf 2007   & A,C \nl
20462$+$1554 & STF & 2725 & AB    & 2016.727 &  11.4 & 0.2 &  6.189 & 0.021 & 1 &  $-$0.3 & \phs0.042 & Mason \& Hartkopf 2014      &   \nl
20467$+$1607 & STF & 2727 & AB    & 2005.485 & 265.9 & 0.2 &  9.061 & 0.031 & 1 & \phs0.3 &  $-$0.100 & Hale 1994                   & A,C \nl
20467$+$1607 & STF & 2727 & AB    & 2016.727 & 265.6 & 0.2 &  8.966 & 0.030 & 1 & \phs0.6 & \phs0.025 & Hale 1994                   &   \nl
20520$+$4346 & STT &  416 & AB    & 2016.853 & 118.1 & 0.2 &  9.606 & 0.033 & 1 & \phs0.5 & \phs0.112 & Hartkopf \& Mason 2011b     & B \nl
21003$+$6130 & HJ  & 1607 & AB    & 2016.700 &  72.8 & 0.2 & 15.729 & 0.053 & 1 &  $-$0.3 & \phs0.073 & Hartkopf \& Mason 2011b     &   \nl
21031$+$0132 & STF & 2744 & AB    & 2016.716 & 109.6 & 0.2 &  1.227 & 0.009 & 2 & \phs6.6 & \phs0.034 & Popovic 1969                & A \nl
21069$+$3845 & STF & 2758 & AB    & 2016.664 & 152.7 & 0.2 & 31.855 & 0.108 & 1 & \phs0.7 & \phs0.205 & Gorshanov et al.\ 2006      &   \nl
21144$+$2905 & STF & 2779 & AB    & 2016.659 & 164.1 & 0.2 & 15.212 & 0.052 & 1 & \phs0.0 & \phs0.064 & Hartkopf \& Mason 2011b     &   \nl
21199$+$5841 & HLM &   39 &       & 2016.661 &  63.3 & 0.2 & 18.771 & 0.064 & 1 & \phs0.5 &  $-$0.014 & Hartkopf \& Mason 2011b     &   \nl
21200$+$5259 & STF & 2789 & AB    & 2016.661 & 113.4 & 0.2 &  6.910 & 0.023 & 1 &  $-$0.5 & \phs0.012 & Kiselev et al.\ 2009        & E \nl
21208$+$3227 & STT &  437 & AB    & 2016.659 &  19.3 & 0.2 &  2.481 & 0.008 & 1 & \phs0.8 & \phs0.042 & Hartkopf \& Mason 2011a     & A \nl
21264$-$2025 & HWE &   56 & AC    & 2016.727 & 160.6 & 0.2 &  5.074 & 0.017 & 1 & \phs0.4 & \phs0.004 & Hartkopf \& Mason 2011b     &   \nl
21289$+$1105 & STF & 2799 & AB    & 2016.653 & 259.4 & 0.2 &  1.911 & 0.006 & 1 & \phs1.1 & \phs0.033 & Hartkopf \& Mason 2011a     & A \nl
21328$+$5839 & HJ  & 1659 & AB    & 2016.785 & 287.3 & 0.2 &  7.808 & 0.027 & 1 &  $-$0.1 &  $-$0.016 & Hartkopf \& Mason 2011b     &   \nl
21330$+$2043 & STF & 2804 & AB    & 2016.653 & 358.5 & 0.2 &  3.375 & 0.011 & 1 & \phs0.0 & \phs0.034 & Hartkopf \& Mason 2011b     & B \nl
21370$+$8255 & STF & 2837 &       & 2016.700 & 269.2 & 0.2 &  3.205 & 0.011 & 1 & \phs0.9 & \phs0.119 & Kiselev et al.\ 2009        & A \nl
21520$+$5548 & STF & 2840 & AB    & 2016.683 & 196.1 & 0.2 & 17.789 & 0.060 & 1 & \phs0.3 & \phs0.079 & Hartkopf \& Mason 2011b     &   \nl
21555$+$5232 & STT &  456 & AC    & 2016.683 & 292.1 & 0.2 & 26.140 & 0.089 & 1 & \phs1.1 &  $-$0.171 & Hartkopf \& Mason 2011b     &   \nl
21582$+$8252 & STF & 2873 & AB    & 2016.700 &  65.9 & 0.2 & 13.960 & 0.047 & 1 &  $-$0.1 & \phs0.218 & Grosheva 2006               &   \nl
             &     &      &       &          &       &     &        &       &   & \phs0.1 & \phs0.221 & Grosheva 2006               &   \nl
21584$+$5245 & ES  & 1015 & AB    & 2016.683 & 239.4 & 0.2 &  8.197 & 0.028 & 1 & \phs0.5 & \phs0.051 & Hartkopf \& Mason 2011b     &   \nl
22033$+$6051 & STF & 2860 & AB    & 2016.700 & 256.9 & 0.2 & 13.420 & 0.046 & 1 & \phs0.1 & \phs0.234 & Hartkopf \& Mason 2011b     & B \nl
22038$+$6438 & STF & 2863 & AB    & 2005.720 & 274.9 & 0.2 &  8.005 & 0.027 & 1 & \phs0.6 &  $-$0.284 & Zeller 1965                 & A,C \nl
22038$+$6438 & STF & 2863 & AB    & 2016.703 & 274.8 & 0.2 &  8.120 & 0.028 & 1 & \phs1.1 &  $-$0.297 & Zeller 1965                 &   \nl
22057$+$2954 & HJ  & 1721 &       & 2016.815 & 264.7 & 0.2 & 12.759 & 0.043 & 1 &  $-$0.2 &  $-$0.007 & Hartkopf \& Mason 2011b     &   \nl
22086$+$5917 & STF & 2872 & BC    & 2016.703 & 296.2 & 0.2 &  0.777 & 0.003 & 1 &  $-$0.6 &  $-$0.025 & Seymour et al.\ 2002        & A \nl
22143$+$1711 & STF & 2877 & AB    & 2016.662 &  24.2 & 0.2 & 23.935 & 0.081 & 1 & \phs0.5 & \phs0.372 & Hartkopf \& Mason 2011b     & B \nl
22207$+$2457 & STF & 2895 & AB    & 2016.757 &  48.3 & 0.2 & 14.001 & 0.048 & 1 & \phs0.3 & \phs0.094 & Hartkopf \& Mason 2011b     &   \nl
22226$+$3328 & ES  &  390 &       & 2016.760 & 268.1 & 0.2 &  8.409 & 0.029 & 1 &  $-$0.4 & \phs0.145 & Hartkopf \& Mason 2011b     & B \nl
22241$-$0450 & BU  &  172 & AB    & 2016.727 &  30.9 & 0.2 &  0.461 & 0.002 & 1 & \phs1.2 &  $-$0.011 & Tokovinin et al.\ 2015      & A \nl
22266$-$1645 & SHJ &  345 & AB    & 2016.727 &  71.9 & 0.2 &  1.206 & 0.004 & 1 & \phs2.6 &  $-$0.078 & Hale 1994                   & A \nl
22288$-$0001 & STF & 2909 & AB    & 2016.727 & 162.6 & 0.2 &  2.343 & 0.008 & 1 & \phs3.8 & \phs0.136 & Tokovinin 2016              & A \nl
22326$+$0725 & STF & 2915 & AB    & 2016.739 & 125.5 & 0.6 & 15.261 & 0.023 & 2 & \phs0.6 & \phs0.079 & Hartkopf \& Mason 2011b     &   \nl
22361$+$7253 & BU  & 1092 & AB    & 2016.700 & 222.6 & 0.2 &  0.358 & 0.001 & 1 & \phs2.9 &  $-$0.005 & Docobo \& Campo 2016        & A \nl
22396$-$1237 & STF & 2928 & AB    & 2016.757 & 280.6 & 0.2 &  3.076 & 0.010 & 1 &  $-$0.6 &  $-$0.118 & Hartkopf \& Mason 2011b     & B \nl
22419$+$2126 & STF & 2934 &       & 2016.910 &  56.5 & 0.2 &  1.401 & 0.005 & 3 & \phs2.8 & \phs0.040 & Zirm 2013                   & A \nl
22460$+$1915 & STF & 2941 &       & 2016.880 & 258.2 & 0.3 & 14.765 & 0.009 & 2 & \phs0.5 & \phs0.039 & Hartkopf \& Mason 2011b     &   \nl
22478$-$0414 & STF & 2944 & AB    & 2016.757 & 305.2 & 0.2 &  1.833 & 0.006 & 1 & \phs0.2 & \phs0.014 & Zirm 2007                   &   \nl
22478$-$0414 & STF & 2944 & AC    & 2016.757 &  85.8 & 0.2 & 62.183 & 0.211 & 1 & \phs0.6 & \phs0.928 & Hartkopf \& Mason 2011b     & B \nl
22490$+$6834 & STF & 2947 & AB    & 2016.850 &  55.7 & 0.2 &  4.681 & 0.016 & 1 & \phs0.6 &  $-$0.013 & Hartkopf \& Mason 2011b     &   \nl
22557$+$1547 & HU  &  987 &       & 2016.911 &  76.2 & 0.1 &  1.196 & 0.004 & 3 &  $-$0.5 & \phs0.046 & Brendley \& Hartkopf 2007   & A \nl
23077$+$0636 & STF & 2976 & AC    & 2016.956 & 209.4 & 0.3 & 21.373 & 0.033 & 2 & \phs1.0 & \phs0.148 & Hartkopf \& Mason 2011b     &   \nl
23114$+$3813 & HO  &  197 & AB,C  & 2016.817 & 319.4 & 0.2 & 36.007 & 0.122 & 1 & \phs0.1 & \phs0.230 & Hartkopf \& Mason 2011b     & B \nl
23114$+$3813 & HO  &  197 & AB,D  & 2016.817 & 278.9 & 0.2 & 57.025 & 0.194 & 1 & \phs0.2 & \phs0.330 & Hartkopf \& Mason 2011b     & B \nl
23133$+$2205 & STF & 2990 & AB    & 2016.760 &  56.0 & 0.2 &  2.598 & 0.009 & 1 & \phs0.6 & \phs0.042 & Hartkopf \& Mason 2011b     & B \nl
23186$+$6807 & STF & 3001 & AB    & 2002.559 & 220.7 & 0.2 &  3.489 & 0.012 & 1 & \phs0.3 & \phs0.187 & Docobo et al.\ 2003         & A \nl
23212$+$3526 & STF & 3006 & AB    & 2016.839 & 151.9 & 0.2 &  7.341 & 0.025 & 1 & \phs0.7 & \phs0.011 & Hartkopf \& Mason 2011b     &   \nl
23238$-$0828 & STF & 3008 &       & 2016.815 & 147.7 & 0.2 &  6.984 & 0.024 & 1 & \phs0.2 & \phs0.092 & Hartkopf \& Mason 2011b     & B \nl
23317$+$1956 & WIR &    1 & AB    & 2004.923 &  86.9 & 0.2 &  5.285 & 0.018 & 2 &  $-$1.3 &  $-$0.148 & Heintz 1984                 & A \nl
23431$+$1150 & A   & 1242 &       & 2016.826 & 339.6 & 0.2 &  1.056 & 0.004 & 1 &  $-$0.1 & \phs0.069 & Ling 2004                   &   \nl
23487$+$6453 & STT &  507 & AB    & 2016.785 & 322.3 & 0.2 &  0.731 & 0.002 & 1 & \phs3.7 & \phs0.023 & Zulevic 1977                & A \nl
23536$+$5131 & STT & A251 & AB    & 2016.785 & 208.3 & 0.2 & 48.242 & 0.164 & 1 & \phs0.5 & \phs0.319 & Hartkopf \& Mason 2011b     &   \nl
23564$-$0930 & STF & 3046 & AB    & 2016.818 & 269.1 & 0.2 &  3.953 & 0.013 & 1 & \phs0.8 &  $-$0.038 & Hartkopf \& Mason 2011b     &   \nl
23568$+$0444 & A   & 2100 &       & 2002.854 & 277.3 & 0.2 &  0.328 & 0.001 & 1 & \phs1.1 & \phs0.051 & Mason  \& Hartkopf 2012     &   \nl
23595$+$3343 & STF & 3050 & AB    & 2016.818 & 340.6 & 0.2 &  2.465 & 0.008 & 1 &  $-$0.0 & \phs0.055 & Hartkopf \& Mason 2011a     & A \nl
\enddata
\tablenotetext{~}{\phm{1}A : Residuals indicate this system should be monitored for possible orbit improvement.}
\tablenotetext{~}{\phm{1}B : Residuals indicate this linear solution may be ready for update soon.}
\tablenotetext{~}{\phm{1}C : This would have been published in an earlier paper in this series, but given the large change in position
                  this measure was not published until its confirmation.}
\tablenotetext{~}{\phm{1}D : Two solutions in Kiselev et al.\ (2009). The one whose residuals are given here ($\Omega = 242^{\circ}$) is
                  definitely better.}
\tablenotetext{~}{\phm{1}E : Two solutions in Kiselev et al.\ (2009). The one whose residuals are given here ($\Omega = 290^{\circ}$) is
                  definitely better.}
\end{deluxetable}


\tiny
\begin{deluxetable}{ll@{~}r@{~}lrrrrrrrlc}
\label{tab:orbit}
\tablenum{6}
\tablewidth{0pt}
\tablecaption{Improved Orbital Elements}
\tablehead{
\colhead{WDS} & 
\multicolumn{3}{c}{Discoverer} &
\colhead{~~~P} & 
\colhead{~a} & 
\colhead{~~i} &
\colhead{~~$\Omega$} &
\colhead{~~T$_0$} &
\colhead{~~e} & 
\colhead{~~$\omega$} &
\colhead{Reference} &
\colhead{Gr} \nl
\colhead{Designation} &   
\multicolumn{3}{c}{Designation} &
\colhead{~~~(yr)} & 
\colhead{~($''$)} & 
\colhead{~~($^{\circ}$)} &
\colhead{~~($^{\circ}$)} &
\colhead{~~(yr)} &
\colhead{ } & 
\colhead{~~($^{\circ}$)}  &
\colhead{ } &
\colhead{ } \nl 
}
\startdata
\multicolumn{13}{c}{Improved but still Provisional Orbital Elements} \nl
\tableline \nl
00162$+$7657 & STF &   13 &       &      1245.6 &      1.078\phn    &     140.7 &      86.1    &      3048.6\phn &      0.472\phn &     287.6    & Olevic \& Jovanovic 2001  & 4 \nl
\tableline \nl
\vspace{-5pt}\nl
\multicolumn{13}{c}{Reliable Orbital Elements} \nl
\tableline \nl
08269$+$3212 & HU  &  714 & Ba,Bb &       195.5 &      0.535\phn    &      72.4 &     135.9    &      1943.2\phn &      0.567\phn &       3.7    & Baize 1993                & 4 \nl
             &     &      &       &   $\pm$13.2 & $\pm$0.015\phn    &  $\pm$3.3 &  $\pm$1.6    &    $\pm$2.5\phn & $\pm$0.040\phn &  $\pm$6.7    &                           &   \nl             
\vspace{-5pt}\nl
19282$-$1209 & SCJ &   22 &       &       170.2 &      1.0031       &      10.4 &     206.\phn &      1983.91    &      0.5837    &     308.\phn & Docobo \& Ling 2009       & 3 \nl
             &     &      &       &    $\pm$1.3 & $\pm$0.0059       &  $\pm$3.0 & $\pm$18.\phn &    $\pm$0.15    & $\pm$0.0038    & $\pm$17.\phn &                           &   \nl             
\vspace{-5pt}\nl
20213$+$0250 & HLD &  158 &       &       232.4 &      0.858\phn    &     128.6 &     218.3    &      1961.16    &      0.746\phn &      17.7    & Heintz 1998               & 4 \nl
             &     &      &       &    $\pm$7.8 & $\pm$0.017\phn    &  $\pm$2.1 &  $\pm$1.9    &    $\pm$0.52    & $\pm$0.012\phn &  $\pm$2.6    &                           &   \nl             
\vspace{-5pt}\nl
\enddata
\end{deluxetable}

\scriptsize

\begin{deluxetable}{ll@{~}r@{~}lrrrrr}
\label{tab:lephem}
\tablenum{7}
\tablewidth{0pt}
\tablecaption{Ephemerides for New Orbit Solutions}
\tablehead{
\colhead{WDS} & 
\multicolumn{3}{c}{Discoverer} &
\multicolumn{1}{c}{2018.0} &
\multicolumn{1}{c}{2020.0} &
\multicolumn{1}{c}{2022.0} &
\multicolumn{1}{c}{2024.0} &
\multicolumn{1}{c}{2026.0} \nl
\colhead{Designation} &   
\multicolumn{3}{c}{Designation} &
\colhead{$\theta$    ~~~~$\rho$} & 
\colhead{$\theta$    ~~~~$\rho$} & 
\colhead{$\theta$    ~~~~$\rho$} & 
\colhead{$\theta$    ~~~~$\rho$} & 
\colhead{$\theta$    ~~~~$\rho$} \nl
\colhead{~} &   
\multicolumn{3}{c}{~} &
\colhead{($\circ$)    ~~($''$)} & 
\colhead{($\circ$)    ~~($''$)} & 
\colhead{($\circ$)    ~~($''$)} & 
\colhead{($\circ$)    ~~($''$)} & 
\colhead{($\circ$)    ~~($''$)}
}
\startdata
00162$+$7657 & STF &   13 &       &  48.3\phn\phn0.957 &  47.8\phn\phn0.960 &  47.3\phn\phn0.962 &  46.8\phn\phn0.964 &  46.3\phn\phn0.967 \nl
08269$+$3212 & HU  &  714 & Ba,Bb & 312.6\phn\phn0.792 & 313.0\phn\phn0.801 & 313.4\phn\phn0.808 & 313.8\phn\phn0.815 & 314.2\phn\phn0.821 \nl
19282$-$1209 & SCJ &   22 &       & 290.9\phn\phn1.131 & 293.5\phn\phn1.166 & 296.0\phn\phn1.199 & 298.3\phn\phn1.231 & 300.4\phn\phn1.261 \nl
20213$+$0250 & HLD &  158 &       &  41.7\phn\phn1.209 &  41.1\phn\phn1.230 &  40.4\phn\phn1.250 &  39.8\phn\phn1.269 &  39.3\phn\phn1.287 \nl
\enddata
\end{deluxetable}


\scriptsize
\begin{deluxetable}{ll@{~}rccclccc}
\tablenum{8}
\tablewidth{0pt}
\tablecaption{Double Stars Not Found}
\tablehead{
\colhead{WDS} & 
\multicolumn{2}{c}{Discoverer} &
\multicolumn{4}{c}{Most Recent Published Observation} &
\multicolumn{2}{c}{Published Magnitudes} &
\colhead{Notes} \nl
\colhead{Designation} &
\multicolumn{2}{c}{Designation} & 
\colhead{Date} &
\colhead{$\theta$} &
\colhead{$\rho$} &
\colhead{Reference} &
\colhead{Primary} &
\colhead{Secondary} &
\colhead{~} \nl
\colhead{~} &
\multicolumn{2}{c}{~} & 
\colhead{~} &
\colhead{(${\circ}$)} & 
\colhead{($''$)} &
\colhead{~} &
\colhead{~} &
\colhead{~} &
\colhead{~}
}
\startdata
03472$+$2522 & POU &  310 & 1894 &    100 &    16.0\phn & Pourteau 1933 & 11.4 & 11.4 &   \nl
07078$-$0822 & OL  &  168 & 1935 &    157 & \phn4.09    & Olivier 1939  & 10.0 & 10.4 &   \nl
08419$+$3546 & SEI &  508 & 1895 & \phn61 &    25.70    & Scheiner 1908 & 11.0 & 11.0 & 1 \nl
23194$+$2417 & POU & 5798 & 1894 & \phn90 &    19.1\phn & Pourteau 1933 & 11.8 & 11.8 &   \nl
\enddata
\tablenotetext{~}{1: Neither component seen on POSS plate; may be flaws on AC Potsdam plate.
Also unresolved in Berko (2009).}
\end{deluxetable}


\begin{references}



\reference {} Aitken, R.G.\ \& Doolittle, E.\ 1932, New General 
              Catalogue of Double Stars Within 120$^{\circ}$ of
              the North Pole (Washington, DC: Carnegie Institution
              of Washington)

\reference {} Alzner, A.\ 1998, A\&AS 132, 253

\reference {} Alzner, A.\ 2007, Inf.\ Circ.\ 163

\reference {} Bagnuolo, W.G., Mason, B.D., Barry, D., Hartkopf, W. 
              \& McAlister, H.A. 1992, AJ 103, 1399

\reference {} Baize, P.\ 1978, Inf.\ Circ.\ 76

\reference {} Baize, P.\ 1993, A\&AS 99, 205

\reference {} Berko, E. 2009, JDSO 5, 168

\reference {} Brendley, M.\ \& Hartkopf, W.I.\ 2007, Inf.\ Circ.\ 
              163

\reference {} Chang, L.\ 1972, AJ 77, 759

\reference {} Cutri, R.M., Skrutskie, M.F., van Dyk, S., et al.\ 
              2003, yCat, 2246, 0

\reference {} Cutri, R.M., et al.\ 2012, yCat, 2311, 0

\reference {} Cvetkovic, Z.\ \& Novakovic, B.\ 2006, Serbian AJ 173,
              73

\reference {} Cvetkovic, Z., Pavlovic, R.\ \& Boeva, S.\ 2016, AJ
              151, 58

\reference {} DENIS consortium 2005, yCat, 2263, 0

\reference {} Docobo, J.A.\ \& Campo, P.\ 2016, Inf.\ Circ.\ 188

\reference {} Docobo, J.A.\ \& Costa, J.M.\ 1984, Inf.\ Circ.\ 92

\reference {} Docobo, J.A.\ \& Ling, J.F.\ 2009, AJ 138, 1159

\reference {} Docobo, J.A., Tamazian, V.S., Andrade, M.\ \&
              Melikian, N.D.\ 2003, AJ 126, 1522

\reference {} Drummond, J.D., Christou, J.C.\ \& Fugate, R.Q.\ 1995,
              ApJ 450, 380

\reference {} Eggenberger, P., Miglio, A., Carrier, F., Fernandes, 
              J.\ \& Santos, N.C.\ 2008, A\&A 482, 631

\reference {} Friedman, E.A., Mason, B.D.\ \& Hartkopf, W.I.\ 2012,
              Inf.\ Circ.\ 176

\reference {} Genet, R., Zirm, H., Rica, F., Richards, J., Rowe, D.\
              \& Gray, D.\ 2015, JDSO 11, 200

\reference {} Gorshanov, D.L., Shakht, N.A.\ \& Kisselev, A.A.\
              2006, SvA 49, 386

\reference {} Grosheva, E.A.\ 2006, SvA 49, 397

\reference {} Hale, A.\ 1994, AJ 107, 306

\reference {} Hartkopf, W.I.\ \& Mason, B.D.\ 2011a, AJ 142, 56

\reference {} Hartkopf, W.I. \& Mason, B.D. 2011b, {\it Catalog of
              Rectilinear Elements}, published in {\it Second USNO
              Double Star CD 2006.5}\footnote{See the current 
              version at
              {\tt http://ad.usno.navy.mil/wds/lin1.html}}.

\reference {} Hartkopf, W.I.\ \& Mason, B.D.\ 2011c, Inf.\ Circ.\
              175

\reference {} Hartkopf, W.I.\ \& Mason, B.D.\ 2015, AJ 150, 136

\reference {} Hartkopf, W.I., Mason, B.D. \& Worley, C.E. 2001, AJ
              122, 3472\footnote{See the current version at 
              {\tt http://ad.usno.navy.mil/wds/orb6.html}.}

\reference {} Hartkopf, W.I., Mason, B.D., Finch, C.T., Zacharias,
              N., Wycoff, G.L.\ \& Hsu, D.\ 2013, AJ 146, 76

\reference {} Hartkopf, W.I., McAlister, H.A. \& Franz, O.G. 1989,
              AJ 98, 1014

\reference {} Heintz, W.D.\ 1984, AJ 89, 1063

\reference {} Heintz, W.D.\ 1996, AJ 111, 408

\reference {} Heintz, W.D.\ 1998, ApJS 111, 335

\reference {} H$\o$g, E., Fabricius, C., Makarov, V.V., Urban, S.,
              Corbin, T., Wycoff, G., Bastian, U., Schwekendiek, P.,
              \& Wicenec, A. 2000, A\&A 357, 367

\reference {} Hopmann, J.\ 1960, Mitt.\ Sternw.\ Wien 10, 155

\reference {} Hopmann, J.\ 1964, Ann.\ Sternw.\ Wien 26, \#1

\reference {} Hopmann, J.\ 1970, Astron.\ Mitt.\ Wien, \#5, 217

\reference {} Hopmann, J.\ 1973, Astron.\ Mitt.\ Wien \#13, 301

\reference {} Hurowitz, J.L., Hartkopf, W.I.\ \& Mason, B.D.\ 2013,
              Inf.\ Circ.\ 181

\reference {} Hurowitz, J.L., Hartkopf, W.I.\ \& Mason, B.D.\ 2014,
              Inf.\ Circ.\ 182

\reference {} Kiselev, A.A.\ \& Romanenko, L.G.\ 1996, Astr.\
              Reports 40, 795

\reference {} Kiselev, A.A., Romanenko, J.G.\ \& Kalinichenko, O.A.\
              2009, Astr.\ Reports 53, 126

\reference {} Kiyaeva, O.V., Gorynya, N.A.\ \& Izmailov, I.S.\ 2010,
              SvAL 36, 204

\reference {} Kiyaeva, O.V., Kisselev, A.A., Polyakov, E.V.\ \&
              Rafal\'skii, V.B.\ 2001, SvAL 27, 391

\reference {} Ling, J.F.\ 2004, Inf.\ Circ.\ 154

\reference {} Mason, B.D.\ \& Hartkopf, W.I.\ 2012, Inf.\ Circ.\ 178

\reference {} Mason, B.D.\ \& Hartkopf, W.I.\ 2014, Inf.\ Circ.\ 184

\reference {} Mason, B.D., Douglass, G.G.\ \& Hartkopf, W.I.\ 1999,
              AJ 117, 1023

\reference {} Mason, B.D., Hartkopf, W.I., Bredthauer, G., Ferguson,
              E.W., Finch, C.T., Killian, C.M., Rafferty, T.J.,
              Ragan, T.J. \& Wieder, G.D. 2017, AJ 153, 20

\reference {} Mason, B.D., Hartkopf, W.I., Wycoff, G.L., Rafferty,
              T.J., Urban, S.E.\ \& Flagg, L.\ 2004, AJ 128, 3012

\reference {} McAlister, H.A., Hartkopf, W.I., Hutter, D.J.\ \&
              Franz, O.G.\ 1987, AJ 93, 688

\reference {} Mason, B.D., Wycoff, G.L., Hartkopf, W.I., Douglass,
              G.G.\ \& Worley, C.E.\ 2001, AJ 122,
              3466\footnote{see current version at 
              {\tt http://ad.usno.navy.mil/wds/}.}

\reference {} Novakovic, B.\ \& Todorovic, N.\ 2006, Serbian AJ 172,
              21

\reference {} Olevic, D.\ \& Jovanovic, P.\ 2001, Serbian AJ 163, 5

\reference {} Olevic, D., Popovic, G.M., Pavlovic, R.\ \& Cvetkovic,
              Z.\ 2003, Serbian AJ 166, 43

\reference {} Olivier, C.P.\ 1939, Pub.\ Univ.\ Penn.\ 5, Pt.\ 2,
              Sec.\ 1

\reference {} Pavlovic, R\ \& Todorovic, N.\ 2005, Serbian AJ 170,
              73

\reference {} Popovic, H.M. 1969, Bull.\ Obs.\ Astron.\ Belgrade 27,
              \#1, 33

\reference {} Popovic, G.M.\ \& Pavlovic, R.\ 1996, Bull.\ Obs.\
              Astron.\ Belgrade \#153, 57

\reference {} Pourteau, M.A.\ 1933, Cat.\ des etoiles doubles de la
              zone $+$24deg de la carte photog.\ du ciel

\reference {} Prieur, J.-L., Scardia, M., Pansecchi, L., Argyle,
              R.W.\ \& Sala, M.\ 2012, MNRAS 422, 1057

\reference {} Raghavan, D., McAlister, H.A., Torres, G., Latham,
              D.W., Mason, B.D., Boyajian, T.S., Baines, E.K.,
              Williams, S.J. ten Brummelaar, T.A., Farrington, C.D.,
              Ridgway, S.T., Sturmann, L., Sturmann, J.\ \& Turner, 
              N.H.\ 2008, ApJ 690, 394

\reference {} Riddle, R.L., Tokovinin, A., Mason, B.D., Hartkopf,
              W.I., Roberts, L.C., Jr., Baranec, C., Law, N.M., Bui,
              K., Burse, M.P., Das, H.K., Dekany, R.G., Kulkarni,
              S., Punnadi, S., Ramaprakash, A.N.\ \& Tendulkar,
              S.P.\ 2015, ApJ 799, 4

\reference {} Rica, F.M. 2012, JDSO 8, 127

\reference {} Rica, F.M., Barrena, R., Vazquez, G., Henriquez, J.A.\
              \& Hernandez, F.\ 2012, MNRAS 419, 197

\reference {} Romanenko, L.G.\ \& Kiselev, A.A.\ 2014, Astron.\
              Reports 58, 30

\reference {} Scardia, M., Prieur, J.-L., Pansecchi, L.\ \& Argyle,
              R.W.\ 2012, Inf.\ Circ.\ 177

\reference {} Scardia, M., Prieur, J.-L., Pansecchi, L., Argyle,
              R.W.\ \& Sala, M.\ 2011, AN 332, 508

\reference {} Scardia, M., Prieur, J.-L., Pansecchi, L., Argyle,
              R.W.\ \& Zanutta, A.\ 2015a, Inf.\ Circ.\ 185

\reference {} Scardia, M., Prieur, J.-L., Pansecchi, L., Argyle,
              R.W., Zanutta, A.\ \& Aristidi, E.\ 2015b, AN 336, 388

\reference {} Scardia, M., Prieur, J.-L., Pansecchi, L., Argyle,
              R.W.\ \& Zanutta, A.\ 2015c, Inf.\ Circ.\ 186

\reference {} Scheiner, J.\ 1908, Pub.\ Obs.\ Potsdam 20, \#59

\reference {} Seymour, D., Mason, B.D., Hartkopf, W.I.\ \& Wycoff,
              G.L.\ 2002, AJ 123, 1023

\reference {} Sinnott, R.W.\ 1999, Sky \& Telescope 97, 100

\reference {} S\"{o}derhjelm, S.\ 1999, A\&A 341, 121

\reference {} Tokovinin, A.\ 2016, ApJ 831, 151

\reference {} Tokovinin, A., Mason, B.D., Hartkopf, W.I., Mendez,
              R.A.\ \& Horch, E.P.\ 2015, AJ 150, 50

\reference {} Urban, S.E., Corbin, T.E., Wycoff, G.L., Martin, J.C.,
              Jackson, E.S., Zacharias, M.I., \& Hall, D.M. 1998, AJ
              115, 1212

\reference {} van Leeuwen, F.\ 2007, A\&A 474, 653

\reference {} Wycoff, G.L., Mason, B.D.\ \& Urban, S.E.\ 2006, AJ
              132, 50

\reference {} Zacharias, N., Finch, C.T., Girard, T.M., Henden, A.,
              Bartlett, J.L., Monet, D.G., \& Zacharias, M.I. 2013,
              AJ 145, 44

\reference {} Zasche, P., Wolf, M., Hartkopf, W.I., Svoboda, P.,
              Uhlar, R., Liakos, A.\ \& Gazeas, K.\ 2009, AJ 138,
              664

\reference {} Zeller, G.\ 1965, Ann.\ Sternw.\ Wien 26, 111

\reference {} Zirm, H.\ 2007, Inf.\ Circ.\ 161

\reference {} Zirm, H.\ 2008, Inf.\ Circ.\ 166

\reference {} Zirm, H.\ 2011, JDSO 7, 24

\reference {} Zirm, H.\ 2013, JDSO 9, 214

\reference {} Zirm, H.\ 2014, Inf.\ Circ.\ 182

\reference {} Zirm, H.\ 2015, Inf.\ Circ.\ 185

\reference {} Zirm, H.\ \& Rica, F.M.\ 2014, Inf.\ Circ.\ 183 

\reference {} Zulevic, D.J.\ 1977, Inf.\ Circ.\ 72

\end{references}
\end{document}